\documentclass[twocolumn,showpacs,amssymb,aps]{revtex4}

\usepackage{graphicx}
\begin{document}

\title{Unruh Effect in the General Light-Front Frame}  

\author{Ashok Das$^{a}$, J. Frenkel$^{b}$ and Silvana Perez$^{c}$}
\affiliation{$^{a}$ Department of Physics and Astronomy,
University of Rochester,
Rochester, New York 14627-0171, USA}
\affiliation{$^{b}$ Instituto de F\'{\i}sica, Universidade de S\~{a}o
Paulo, S\~{a}o Paulo, BRAZIL} 
\affiliation{$^{c}$ Departamento de F\'{\i}sica, 
Universidade Federal do Par\'{a}, 
Bel\'{e}m, Par\'{a} 66075-110, BRAZIL}

\bigskip

\begin{abstract}

We study the phenomenon of Unruh effect in a massless scalar field theory
quantized on the light-front in the general light-front frame. We
determine the uniformly accelerating coordinates in such a frame and
through a direct transformation show that the propagator of the theory
has a thermal character in the  uniformly accelerating coordinate
system with a temperature given by Tolman's law. We also carry out a
systematic analysis of this phenomenon from the Hilbert space point of
view and show that the vacuum of this theory appears as a thermal
vacuum to a Rindler observer with the same temperature as given by
Tolman's law.

\end{abstract}


\maketitle

\section{Introduction}

It has been observed in recent years that a statistical description of
theories quantized on the light-front \cite{dirac,brodsky} prefers a
general coordinate frame \cite{das,weldon,das1}. Denoting the
Minkowski coordinates by $x^{\mu} = (t,x,y,z)$ and the coordinates of
the general light-front frame by $\bar{x}^{\mu}
= (\bar{t},\bar{x},\bar{y},\bar{z})$, the relation between the two can
be written as \cite{perez} 
\begin{eqnarray}
& & \bar{t} = t + z,\quad \bar{z} = A t + B z,\nonumber\\
& & \bar{x} = x,\qquad\  \bar{y} = y,\label{generalcoordinate}
\end{eqnarray}
where $A,B$ are arbitrary constants with the restriction that $|B|\geq
|A|$ which arises if we require $\bar{t}$ to correspond to the time
variable. The metric in the general light-front frame (GLF) has the
form (We refer the reader to \cite{perez} for details on notations as
well as various other properties of GLF.)
\begin{equation}
\bar{g}^{\rm (GLF)}_{\mu\nu} = \left(\begin{array}{crrc}
\frac{A+B}{B-A} & 0 & 0 & -\frac{1}{B-A}\\
0 & -1 & 0 & 0\\
0 & 0 & -1 & 0\\
- \frac{1}{B-A} & 0 & 0 & 0
\end{array}\right),\label{metric}
\end{equation}
so that the line element is given by 
\begin{equation}
\mathrm{d}\tau^{2} = \frac{A+B}{B-A}\ \mathrm{d}\bar{t}^{2} -
\mathrm{d}\bar{x}^{2} - \mathrm{d}\bar{y}^{2} - \frac{2}{B-A}\
\mathrm{d}\bar{t}\mathrm{d}\bar{z}.\label{lineelement}
\end{equation}

For $A=-B=1$, eq. (\ref{generalcoordinate}) represents the
conventional light-front frame (CLF) while for $A=0, B=1$ we have the
oblique light-front frame (OLF) where most of the discussions of
statistical mechanics have been carried out thus far
\cite{das,weldon,das1,brodsky1,beyer}. In general, if
a quantum field theory quantized at equal time $t$ is at temperature
$T_{\rm M}$ in the Minkowski space, then by Tolman's law \cite{tolman}, the
corresponding  temperature for the theory quantized at equal $\bar{t}$
in the generalized light-front frame is given by
\begin{eqnarray}
\frac{T_{\rm GLF}}{\sqrt{\bar{g}^{\rm (GLF)}_{00}}} & = & \frac{T_{\rm
  M}}{\sqrt{\eta_{00}}} = T_{\rm M}\nonumber\\ 
{\rm or,}\quad T_{\rm GLF} & = & \sqrt{\bar{g}^{\rm (GLF)}_{00}}\ T_{\rm M}
  = \sqrt{\frac{A+B}{B-A}}\ T_{\rm M}.\label{tolman}
\end{eqnarray}
This shows that in the conventional light-front frame where $A=-B=1$
\begin{equation}
T_{\rm CLF} = 0,
\end{equation}
for any finite $T_{\rm M}$ so that any finite temperature in the
Minkowski frame is mapped to zero temperature in the conventional
light-front frame and a
statistical description is not possible. On the other hand, in
the oblique light-front frame where $A=0,B=1$,
\begin{equation}
T_{\rm OLF} = T_{\rm M},
\end{equation}
and the temperature coincides with that corresponding to conventional
quantization. As is clear from (\ref{tolman}), statistical mechanical
description for a theory  quantized on the light-front is possible
as long as $|B|\neq |A|$. In the general light-front frame, the
temperature will be related to that in the Minkowski frame simply
through a scale factor.

It has also been understood for sometime now that equal time quantum
field theories exhibit Unruh effect \cite{unruh} when viewed from a
uniformly accelerating 
coordinate frame. More specifically, a uniformly accelerating observer
sees the 
vacuum of the equal time quantum theory to correspond to a thermal
vacuum with temperature given by (We note that because of the isotropy
of Minkowski space, the direction of acceleration is not important.) 
\begin{equation}
T_{\rm M} = \frac{\alpha}{2\pi},\label{unruhT}
\end{equation}
where $\alpha$ represents the constant proper acceleration of the
observer. It is interesting to study the phenomenon of Unruh effect
within the context of quantum field theories quantized on the
light-front (equal $\bar{t}$) in GLF \cite{perez} for the following
reasons. Since the statistical density matrix in light-front field
theories does not correspond to a naive generalization of the known
density matrix \cite{das,weldon,das1}, further support for the
structure of this density matrix in GLF can be obtained from studying
the thermal behavior of the vacuum in an accelerated coordinate
system. This, of course, immediately raises this interesting issue,
namely, since the form of the line element in (\ref{lineelement}) shows
that  there
are now two distinct possibilities for acceleration (unlike the
Minkowski frame), it is not {\em a priori} clear
whether this would lead to two distinct temperatures for the Unruh
effect (corresponding to the two directions for acceleration) and how
this will be compatible with the unique temperature of the GLF
description following from Tolman's law in (\ref{tolman}).

In this paper, we study these issues systematically. In section {\bf
  II}, we work out the uniformly accelerating coordinates for the two
  cases of acceleration along the $\bar{x}$ axis and the $\bar{z}$
  axis. In section {\bf III}, we show that even though the uniformly
  accelerating coordinates are different for the two cases, the zero
  temperature propagator of a massless scalar field theory quantized
  on the light-front in GLF corresponds to a thermal propagator with
  the unique temperature given by (\ref{tolman}) when transformed to
  the accelerating coordinates \cite{milonni}. In section {\bf IV} we
  carry out the 
  Hilbert space analysis for this theory systematically and show that
  a Rindler observer (uniformly accelerating along $\bar{z}$) perceives the
  GLF vacuum of the theory as a thermal vacuum with the same
  temperature as in (\ref{tolman}). We present a brief summary in
  section {\bf V}.

\section{Uniformly Accelerating Coordinates}

Let us recall that the line element in the general light-front frame has
the form
\begin{equation}
\mathrm{d}\tau^{2} = \frac{A+B}{B-A}\ \mathrm{d}\bar{t}^{2} -
\mathrm{d}\bar{x}^{2} - \mathrm{d}\bar{y}^{2} - \frac{2}{B-A}\
\mathrm{d}\bar{t}\mathrm{d}\bar{z}.\label{lineelement1}
\end{equation}
In this case, the Lorentz contraction factor, in general, has
the form
\begin{equation}
\frac{\mathrm{d}\bar{t}}{\mathrm{d}\tau} = \bar{\gamma} =
\frac{1}{\sqrt{\frac{A+B}{B-A} - \bar{v}_{\bar{x}}^{2} -
    \bar{v}_{\bar{y}}^{2} - \frac{2}{B-A}
    \bar{v}_{\bar{z}}}},\label{contraction}
\end{equation}
with appropriate restrictions on the velocity components. Here the
coordinate velocities are defined as
\begin{equation}
\bar{v}_{\bar{x}} = \frac{\mathrm{d}\bar{x}}{\mathrm{d}\bar{t}},\quad
\bar{v}_{\bar{y}} = \frac{\mathrm{d}\bar{y}}{\mathrm{d}\bar{t}},\quad
\bar{v}_{\bar{z}} = \frac{\mathrm{d}\bar{z}}{\mathrm{d}\bar{t}}.
\end{equation}
From (\ref{lineelement1}) we note that there is an obvious symmetry
between the coordinates 
$\bar{x},\bar{y}$. Therefore, in studying acceleration in such a
frame, there are two distinct cases to consider.

\subsection{Motion along $\bar{x}$}

Let us first consider the case where a particle is moving (and being
uniformly accelerated) along the $\bar{x}$ axis
\cite{rindler}. Therefore, neglecting the
$\bar{y}$ and $\bar{z}$ coordinates, we can define the four velocity
of the particle as
\begin{equation}
\bar{u}^{\mu} = \bar{\gamma} \left(1, \bar{v}_{\bar{x}},0,0\right),
\end{equation}
where we have identified (see (\ref{contraction})) 
\begin{equation}
\bar{v}_{\bar{y}} = 0 = \bar{v}_{\bar{z}},\quad
\bar{\gamma} = \frac{1}{\sqrt{\frac{A+B}{B-A} - \bar{v}_{\bar{x}}^{2}}}.
\end{equation}
Defining the proper acceleration as
\begin{equation}
\bar{a}^{\mu} = \frac{\mathrm{d}\bar{u}^{\mu}}{\mathrm{d}\tau} =
\frac{\mathrm{d}\bar{t}}{\mathrm{d}\tau}
\frac{\mathrm{d}\bar{u}^{\mu}}{\mathrm{d}\bar{t}} = 
\bar{\gamma} \frac{\mathrm{d}\bar{u}^{\mu}}{\mathrm{d}\bar{t}},
\end{equation}
we obtain after some algebra
\begin{equation}
\bar{a}^{\mu} = \left(\bar{v}_{\bar{x}}, \frac{A+B}{B-A}, 0,
0\right)\bar{\gamma}^{4} \frac{\mathrm{d}\bar{v}_{\bar{x}}}{\mathrm{d}\bar{t}}.
\end{equation}
It follows from this that
\begin{equation}
\bar{a}^{2} = \bar{g}^{\rm (GLF)}_{\mu\nu} \bar{a}^{\mu}\bar{a}^{\nu} = -
\left(\sqrt{\frac{A+B}{B-A}}\
\bar{\gamma}^{3}\frac{\mathrm{d}\bar{v}_{\bar{x}}}{\mathrm{d}\bar{t}}
\right)^{2} = - \alpha^{2},
\end{equation}
where $\alpha$ is known as the proper acceleration. In terms of the
proper acceleration $\alpha$, we can write
\begin{equation}
\bar{a}^{\mu} = \frac{\mathrm{d}\bar{u}^{\mu}}{\mathrm{d}\tau} =
\left(\sqrt{\frac{B-A}{A+B}} \bar{\gamma}\bar{v}_{\bar{x}},
\sqrt{\frac{A+B}{B-A}} \bar{\gamma},0,0\right)\alpha.
\end{equation}

For constant $\alpha$, we can now solve the dynamical equations
\begin{equation}
\frac{\mathrm{d}^{2}\bar{x}^{\mu}}{\mathrm{d}\tau^{2}} =
\frac{\mathrm{d}\bar{u}^{\mu}}{\mathrm{d}\tau} = \bar{a}^{\mu},
\end{equation}
to determine the trajectory
\begin{eqnarray}
\bar{\gamma} (\tau) & = & \sqrt{\frac{B-A}{A+B}}\ \cosh
\alpha\tau,\nonumber\\
\bar{v}_{\bar{x}} (\tau) & = & \sqrt{\frac{A+B}{B-A}}\ \tanh
\alpha\tau,\nonumber\\
\bar{t} (\tau) & = & \sqrt{\frac{B-A}{A+B}}\ \frac{1}{\alpha}\ \sinh
\alpha\tau,\nonumber\\
\bar{x} (\tau) & = & \frac{1}{\alpha}\ \cosh
\alpha\tau,\label{xaccelerating} 
\end{eqnarray}
corresponding to the initial conditions $\bar{v}_{\bar{x}} (\tau =0) = 0 =
\bar{t} (\tau=0), \bar{x} (\tau=0) = \frac{1}{\alpha}$.  
It follows now that the trajectory with a constant proper acceleration
defines the hyperbola
\begin{equation}
\bar{x}^{2} = \bar{g}^{\rm (GLF)}_{\mu\nu} \bar{x}^{\mu}\bar{x}^{\nu}
= - \frac{1}{\alpha^{2}},
\end{equation}
with (\ref{xaccelerating}) providing the uniformly accelerating
coordinates for 
the present case. These are quite similar to the case of uniformly
accelerating coordinates in the Minkowski frame except for normalization
factors. 

\subsection{Motion along $\bar{z}$}

Let us consider next the case where the particle is moving (and being
uniformly accelerated) along the $\bar{z}$ axis. In this case, setting
$\bar{x}=\bar{y}=0$, we can write
\begin{equation}
\bar{u}^{\mu} = \frac{\mathrm{d}\bar{x}^{\mu}}{\mathrm{d}\tau} =
\left(\bar{\gamma}, 0, 0, \bar{\gamma}\bar{v}_{\bar{z}}\right),
\end{equation}
where, in the present case, (see (\ref{contraction}))
\begin{equation}
\bar{v}_{\bar{x}} = 0 = \bar{v}_{\bar{y}},\quad
\bar{\gamma} = \sqrt{\frac{B-A}{A+B - 2\bar{v}_{\bar{z}}}}.
\end{equation}
In this case, the acceleration
\begin{equation}
\bar{a}^{\mu} = \frac{\mathrm{d}\bar{u}^{\mu}}{\mathrm{d}\tau},
\end{equation}
can be shown, with a little bit of algebra to have the form
\begin{equation}
\bar{a}^{\mu} = \left(1, 0, 0, (A+B-\bar{v}_{\bar{z}})\right) \bar{\gamma}
\frac{\mathrm{d}\bar{\gamma}}{\mathrm{d}\bar{t}},
\end{equation}
which leads to
\begin{equation}
\bar{a}^{2} = \bar{g}^{\rm (GLF)}_{\mu\nu} \bar{a}^{\mu} \bar{a}^{\nu}
= - \left(\frac{\mathrm{d}\bar{\gamma}}{\mathrm{d}\bar{t}}\right)^{2}
= - \alpha^{2},
\end{equation}
with $\alpha$ representing the proper acceleration. In terms of this,
we can write
\begin{equation}
\bar{a}^{\mu} = \left(1, 0, 0, (A+B-\bar{v}_{\bar{z}})\right)
\bar{\gamma}\alpha.
\end{equation}

For constant $\alpha$, we can solve the dynamical equations
\begin{equation}
\frac{\mathrm{d}^{2}\bar{x}^{\mu}}{\mathrm{d}\tau^{2}} =
\frac{\mathrm{d}\bar{u}^{\mu}}{\mathrm{d}\tau} = \bar{a}^{\mu},
\end{equation}
to obtain the trajectory
\begin{eqnarray}
\bar{\gamma} (\tau) & = & \sqrt{\frac{B-A}{A+B}}\
e^{\alpha\tau},\nonumber\\
\bar{v}_{\bar{z}} (\tau) & = & \frac{A+B}{2}\left(1 -
e^{-2\alpha\tau}\right),\nonumber\\
\bar{t} (\tau) & = & \sqrt{\frac{B-A}{A+B}}\ \frac{1}{\alpha}\
e^{\alpha\tau},\nonumber\\
\bar{z} (\tau) & = & {\rm sgn} (B-A)
\frac{\sqrt{B^{2}-A^{2}}}{\alpha}\ \cosh
\alpha\tau,\label{zaccelerating} 
\end{eqnarray}
where, for simplicity, we have assumed the initial conditions
$\bar{v}_{\bar{z}} (\tau=0) = 0, \bar{t} (\tau=0) =
\sqrt{\frac{B-A}{A+B}} 
\frac{1}{\alpha}$,  and $\bar{z} (\tau=0) = {\rm sgn} (B-A) 
\frac{\sqrt{B^{2}-A^{2}}}{\alpha}$.  (We note here, for later use,
that when $|B|>
|A|$, it follows that ${\rm sgn} (B-A) = {\rm sgn} (A+B)$.) It is
straightforward  to check
that the trajectory (\ref{zaccelerating}) with a constant proper
acceleration defines the hyperbola
\begin{equation}
\bar{x}^{2} = \bar{g}^{\rm (GLF)}_{\mu\nu} \bar{x}^{\mu}\bar{x}^{\nu}
= - \frac{1}{\alpha^{2}},
\end{equation}
and (\ref{zaccelerating}) defines the uniformly  accelerating
coordinates in the present case.

\section{Transformation of the Green's Function}

Let us next consider a massless scalar field theory quantized on the
light-front (at equal $\bar{t}$) in the general light-front frame. In
this case, the field expansion takes the form (we refer the reader to
\cite{perez} for details)
\begin{equation}
\phi (\bar{x}) = \int
\frac{\mathrm{d}^{2}\bar{k}_{\perp}}{(2\pi)^{3/2}}\!\! \int_{0}^{\infty}\!
\frac{\mathrm{d}\bar{k}_{3}}{2\bar{k}_{3}}\!
\left(e^{-i\tilde{\bar{k}}\cdot \bar{x}}\ a (\bar{k}) +
e^{i\tilde{\bar{k}}\cdot \bar{x}}\ a^{\dagger}
(\bar{k})\right)\!,\label{expansion}
\end{equation}
where $\bar{k}_{\perp}$ denotes the transverse components of the
momenta (namely, $\bar{k}_{1},\bar{k}_{2}$) and we have identified
($i=1,2$) 
\begin{equation}
\tilde{\bar{k}}_{\mu} = \left(\bar{k}_{0}, {\rm sgn}
(A-B)\bar{k}_{i}, {\rm sgn} (A-B) \bar{k}_{3}\right),
\end{equation}
with 
\begin{equation}
\bar{k}_{0} = \bar{\omega} = \frac{\bar{k}_{i}^{2} + (B^{2}-A^{2})
  \bar{k}_{3}^{2}}{2|A-B|\bar{k}_{3}} >0,\label{k0}
\end{equation}
for $\bar{k}_{3}>0$. When quantized on the light-front, the creation
and the annihilation operators satisfy the commutation relation
\begin{equation}
\left[a (\bar{k}) , a^{\dagger} (\bar{k}^{\prime})\right] = 2
\bar{k}_{3} \delta^{3} (\bar{k} -
\bar{k}^{\prime}).\label{commutation} 
\end{equation}
The Feynman propagator in the momentum space has the form
\begin{equation}
iD (\bar{k}) = \lim_{\epsilon\rightarrow 0}\  \frac{i}{\bar{k}^{2} +
  i\epsilon},
\end{equation}
while in the coordinate space it takes the form \cite{typo}
\begin{eqnarray}
\langle \phi (\bar{x}_{1})\phi (\bar{x}_{2})\rangle & = & \frac{1}{(2\pi)^{3}}
\int \mathrm{d}^{2}\bar{k}_{\perp} \int_{0}^{\infty}
\frac{\mathrm{d}\bar{k}_{3}}{2\bar{k}_{3}}\ e^{-i\tilde{\bar{k}}\cdot
  (\bar{x}_{1}- \bar{x}_{2})}\nonumber\\
& = & \lim_{\epsilon\rightarrow 0} -
\frac{1}{(2\pi)^{2}}\
\frac{1}{(\bar{x}_{1}-\bar{x}_{2})^{2}-i\epsilon},\label{zeroTpropagator}
\end{eqnarray}
where we are assuming that $\bar{x}_{1}^{0} - \bar{x}_{2}^{0} > 0$.

In the rest frame of the heat bath, the propagator for this theory at
a temperature $T$ can be obtained easily in the momentum space both in
the imaginary time formalism as well as in the real time formalism. In
the imaginary time formalism \cite{kapusta,das2}, it has the form
\begin{equation}
D^{(T_{\rm GLF})} (\bar{k}) = \frac{1}{2(A-B) \bar{k}_{0}\bar{k}_{3} -
  \bar{k}_{\perp}^{2} - (B^{2}-A^{2}) \bar{k}_{3}^{2}},
\end{equation}
with $\bar{k}_{0} = 2i\pi nT_{\rm GLF}$ where $n$ represents an
integer. In the real time formalism
\cite{das2}, on the
other hand, the $++$ component of the propagator can be written as
\begin{equation}
iD_{++}^{(T_{\rm GLF})} (\bar{k}) = \lim_{\epsilon\rightarrow 0}\ 
\frac{i}{\bar{k}^{2}} + 2\pi n(|\bar{k}_{0}|) \delta (\bar{k}^{2}),
\end{equation}
where $n(|k_{0}|)$ represents the Bose-Einstein distribution function
(the Boltzmann constant is assumed to be unity)
\begin{equation}
n(|\bar{k}_{0}|) = \frac{1}{e^{\frac{|\bar{k}_{0}|}{T_{\rm GLF}}} - 1}.
\end{equation}
Here $T_{\rm GLF}$ denotes the temperature in the general light-front 
frame. We can take the Fourier transform of the finite temperature
propagator in either the imaginary time formalism or the real time
formalism with respect to the time variable to obtain the coordinate
space representation of the thermal propagator (which we have
calculated).  Alternatively, we can simply 
evaluate this directly from the field expansion given in
(\ref{expansion}). For a theory quantized on the light-front (see
(\ref{commutation})), we note that
\begin{eqnarray}
\langle a^{\dagger} (\bar{k}) a (\bar{k})\rangle_{T_{\rm GLF}} & = & 2
\bar{k}_{3}\  n (\bar{k}_{0}),\\
\langle a(\bar{k}) a^{\dagger} (\bar{k})\rangle_{T_{\rm GLF}} & = &
2\bar{k}_{3} (1 + n (\bar{k}_{0})),
\end{eqnarray}
where $\bar{k}_{0}$ is defined in (\ref{k0}). Recalling that we are in
the rest frame, we have
\begin{eqnarray}
\langle \phi(\bar{t}_{1}) \phi(\bar{t}_{2})\rangle_{T_{\rm GLF}} & = &\!
\int\! \frac{\mathrm{d}^{2}\bar{k}_{\perp}}{(2\pi)^{3}}\int_{0}^{\infty}\!
\frac{\mathrm{d}\bar{k}_{3}}{2\bar{k}_{3}}\!\left(n (\bar{k}_{0})
e^{i\bar{k}_{0} (\bar{t}_{1}-\bar{t}_{2})}\right.\nonumber\\
 &  & \quad \left.+ (1+n (\bar{k}_{0})) e^{-i\bar{k}_{0}
  (\bar{t}_{1}-\bar{t}_{2})}\right),
\end{eqnarray}
where we are assuming that $\bar{t}_{1}-\bar{t}_{2}>0$.
The integral can be easily done using standard tables \cite{GR} and
recalling 
that in the rest frame, the proper time is related to the coordinate
time as (see (\ref{lineelement1}))
\begin{equation}
\bar{t}_{1} - \bar{t}_{2} = \sqrt{\frac{B-A}{A+B}}\
(\tau_{1}-\tau_{2}) = \sqrt{\frac{B-A}{A+B}}\ \tau,
\end{equation}
we obtain the coordinate representation of the thermal propagator to
be
\begin{eqnarray}
\langle \phi(\bar{\tau}_{1})\phi(\bar{\tau}_{2})\rangle_{T_{\rm GLF}} &
= &  -
\frac{1}{(2\pi)^{2}} \left(\pi T_{\rm
  GLF}\sqrt{\frac{B-A}{A+B}}\right)^{2}\nonumber\\
 &\times&\!\! {\rm cosech}^{2}\!\left(\!\pi T_{\rm
  GLF}\sqrt{\frac{B-A}{A+B}} \tau\right).\label{thermalpropagator}
\end{eqnarray}

Let us next look at the zero temperature propagator
(\ref{zeroTpropagator}) in the accelerating coordinate system. As we
have seen, there are two cases to consider. When the motion is along
the $\bar{x}$ axis, we can set $\bar{y}_{1}=\bar{y}_{2},
\bar{z}_{1}=\bar{z}_{2}$. Furthermore, using the accelerating
coordinate system in (\ref{xaccelerating}), we can write
\begin{eqnarray}
\bar{t}_{1}-\bar{t}_{2} & = & \sqrt{\frac{B-A}{A+B}}\
\frac{1}{\alpha}\left(\sinh \alpha\tau_{1} - \sinh
\alpha\tau_{2}\right),\nonumber\\
\bar{x}_{1} - \bar{x}_{2} & = & \frac{1}{\alpha}\left(\cosh
\alpha\tau_{1} - \cosh \alpha\tau_{2}\right),
\end{eqnarray}
where we are assuming that $\tau_{1}-\tau_{2} = \tau >0$. It follows
now that the invariant length is given by 
\begin{eqnarray}
(\bar{x}_{1}-\bar{x}_{2})^{\mu}(\bar{x}_{1} - \bar{x}_{2})_{\mu} & = &
 \frac{A+B}{B-A} (\bar{t}_{1} -
  \bar{t}_{2})^{2} - (\bar{x}_{1} - \bar{x}_{2})^{2}\nonumber\\
 & = & \frac{4}{\alpha^{2}}\ \sinh^{2} \frac{\alpha\tau}{2}.
\end{eqnarray}
As a result, in the coordinate system (\ref{xaccelerating}) uniformly
accelerating along $\bar{x}$, the zero temperature propagator
(\ref{zeroTpropagator}) takes the form
\begin{equation}
\langle\phi (\bar{x}_{1})\phi(\bar{x}_{2})\rangle = -
\frac{1}{(2\pi)^{2}}\left(\frac{\alpha}{2}\right)^{2} {\rm cosech}^{2}
\frac{\alpha\tau}{2}. 
\end{equation}
Comparing with (\ref{thermalpropagator}), we conclude that an observer
in the frame accelerating along the $\bar{x}$ axis sees the propagator
of the light-front field
theory as a thermal propagator with temperature
\begin{eqnarray}
\frac{\alpha}{2} & = & \pi T_{\rm GLF}
\sqrt{\frac{B-A}{A+B}}\nonumber\\
{\rm or,}\quad T_{\rm GLF} & = & \sqrt{\frac{A+B}{B-A}}\
\frac{\alpha}{2\pi} = \sqrt{\bar{g}^{\rm (GLF)}_{00}}\
\frac{\alpha}{2\pi}. 
\end{eqnarray}

The other case to consider is when the motion is along the $\bar{z}$
axis. In this case, we can set $\bar{x}_{1}=\bar{x}_{2},
\bar{y}_{1}=\bar{y}_{2}$. In the uniformly accelerating coordinates
(\ref{zaccelerating}), we note that we can write 
\begin{eqnarray}
\bar{t}_{1} - \bar{t}_{2} & = & \sqrt{\frac{B-A}{A+B}}\
\frac{1}{\alpha}\left(e^{\alpha\tau_{1}} -
e^{\alpha\tau_{2}}\right),\nonumber\\
\bar{z}_{1}-\bar{z}_{2} & = & {\rm sgn}
(B-A)\frac{\sqrt{B^{2}-A^{2}}}{\alpha}\left(\cosh \alpha\tau_{1} - \cosh
\alpha\tau_{2}\right),\nonumber\\
 &  & 
\end{eqnarray}
where we are again assuming that  $\tau_{1}-\tau_{2} = \tau >0$. It
follows now  that
\begin{eqnarray}
(\bar{x}_{1}-\bar{x}_{2})^{\mu}(\bar{x}_{1}-\bar{x}_{2})_{\mu} & = &
  \frac{A+B}{B-A}
  (\bar{t}_{1}-\bar{t}_{2})^{2}\nonumber\\
&  &  - \frac{2}{B-A}
  (\bar{t}_{1}-\bar{t}_{2})(\bar{z}_{1} - \bar{z}_{2})\nonumber\\
 & = & \frac{4}{\alpha^{2}}\ \sinh^{2} \frac{\alpha\tau}{2},
\end{eqnarray}
so that in the uniformly accelerating coordinates, the zero temperature
propagator (\ref{zeroTpropagator}) takes the form
\begin{equation}
\langle\phi (\bar{x}_{1})\phi(\bar{x}_{2})\rangle = -
\frac{1}{(2\pi)^{2}}\left(\frac{\alpha}{2}\right)^{2} {\rm cosech}^{2}
\frac{\alpha\tau}{2}. 
\end{equation}
Comparing with (\ref{thermalpropagator}), we conclude that an observer
in the frame accelerating along the $\bar{z}$ axis sees the propagator
of the light-front field
theory as a thermal propagator with temperature
\begin{eqnarray}
\frac{\alpha}{2} & = & \pi T_{\rm GLF}
\sqrt{\frac{B-A}{A+B}}\nonumber\\
{\rm or,}\quad T_{\rm GLF} & = & \sqrt{\frac{A+B}{B-A}}\
\frac{\alpha}{2\pi} = \sqrt{\bar{g}^{\rm (GLF)}_{00}}\
\frac{\alpha}{2\pi}.
\end{eqnarray}
In other words, even though there are two distinct possibilities for
acceleration in the GLF, both cases lead to a thermal character for
the Green's function with the same temperature. Furthermore, 
recalling that the Unruh effect predicts $T_{\rm M} =
\frac{\alpha}{2\pi}$ for a conventionally quantized scalar field
theory in Minkowski space (\ref{unruhT}), we recover (\ref{tolman})
in both the cases.

\section{Hilbert Space Analysis}

In the last section, we carried out a very simple analysis where we 
transformed the propagator of a massless scalar field quantized on the
light-front to a uniformly accelerating coordinate system and
thereby showed that it behaves like a thermal propagator with 
temperature $T_{\rm GLF} = \sqrt{\bar{g}_{00}^{\rm (GLF)}}
\frac{\alpha}{2\pi}$ independent of whether the acceleration is along
the $\bar{x}$ ($\bar{y}$) axis or along the $\bar{z}$ axis. This
simple analysis, however, does not bring out many important aspects of
the Hilbert space structure of the theory in the present case which we
will like to investigate systematically in this section. We will do this
only for the case where the acceleration is along the $\bar{z}$ axis
for simplicity. A parallel analysis for the case where the
acceleration is along the $\bar{x}$ axis can be carried out exactly
along the lines to be discussed in this section and does not lead to
any new information. The Hilbert space analysis shows that a Rindler
observer would perceive the vacuum of the theory to correspond to a
thermal vacuum with temperature given in (\ref{tolman}). Such a result
is much more powerful in showing that any matrix elelment of the
theory would appear as a thermal amplitude to an observer in the
accelerating  frame.

Let us first define various relevant coordinate systems associated
with this problem. First, we note that if we define new 
coordinates as ($\bar{x} = \bar{y} = 0$)
\begin{eqnarray}
\bar{t} & = & \sqrt{\frac{B-A}{A+B}}\ X e^{T},\nonumber\\
\bar{z} & = & {\rm sgn} (B-A) 
\sqrt{B^{2}-A^{2}}\ X \cosh T,\label{rindler}
\end{eqnarray}
then, the line element (\ref{lineelement1}) can be written as
\begin{equation}
\mathrm{d}\tau^{2} = X^{2} \mathrm{d}T^{2} -
\mathrm{d}X^{2}.\label{rindlerlineelement}
\end{equation}
Here $X,T$ define the Rindler coordinates \cite{rindler} for the present
case and we have 
\begin{equation}
\bar{x}^{2} = - X^{2},
\end{equation}
so that for constant $X$, they define the hyperbola of constant
acceleration $\alpha = \frac{1}{X}$.

The null geodesics for the theory defined on the general light-front
frame are given (in the present case) by
\begin{equation}
\bar{t} = {\rm constant},\quad \bar{t} - \frac{2}{A+B} \bar{z} = {\rm
  constant},
\end{equation}
which, in turn, allow us to define the null coordinates 
\begin{eqnarray}
u & = & \bar{t}-\frac{2}{A+B} \bar{z} = - \sqrt{\frac{B-A}{A+B}}\ X
e^{-T}\nonumber\\
 & = & - \sqrt{\frac{B-A}{A+B}}\ \frac{1}{a}\ e^{-a (\eta - \xi)} = -
\sqrt{\frac{B-A}{A+B}}\ \frac{1}{a}\ e^{-aU},\nonumber\\
v & = & \bar{t} = \sqrt{\frac{B-A}{A+B}}\ X e^{T}\nonumber\\
 & = & \sqrt{\frac{B-A}{A+B}}\ \frac{1}{a}\ e^{a (\eta + \xi)} =
\sqrt{\frac{B-A}{A+B}}\ \frac{1}{a}\
e^{aV},\label{conformalcoordinates}
\end{eqnarray}
where we have identified \cite{birrell}
\begin{eqnarray}
T & = & a\eta,\qquad X = \frac{e^{a\xi}}{a},\nonumber\\
U & = & \eta - \xi,\quad V = \eta + \xi,\label{conformalcoordinates1}
\end{eqnarray}
with $a$ representing an arbitrary positive constant. The line element
(\ref{lineelement1}) or (\ref{rindlerlineelement}) can now be written
as
\begin{equation}
\mathrm{d}\tau^{2} = e^{2a\xi}\left(\mathrm{d}\eta^{2} -
\mathrm{d}\xi^{2}\right),\label{conformalelement}
\end{equation}
which makes it clear that $\eta,\xi$ define the conformal coordinates
for the system and that
\begin{equation}
\bar{g}^{\rm (conformal)}_{\eta\eta} = - \bar{g}^{\rm
  (conformal)}_{\xi\xi} = e^{2a\xi}.\label{conformalmetric}
\end{equation}
Furthermore, in these coordinates, we have
\begin{equation}
\bar{x}^{2} = - \left(\frac{e^{a\xi}}{a}\right)^{2},
\end{equation}
so that for a constant $\xi$, we have the hyperbola corresponding to
acceleration
\begin{equation}
\alpha = a e^{-a\xi}.\label{acceleration}
\end{equation}

The Rindler wedges, in the present case, are defined by
\begin{equation}
{\rm R}:\quad 0\leq \bar{t}\leq \frac{2}{A+B} \bar{z},\qquad {\rm
  L}:\quad \frac{2}{A+B} \bar{z} \leq \bar{t}\leq 0.\label{wedges}
\end{equation}
In the wedge labeled ``R'', we have (as we have pointed out earlier
${\rm sgn} (A+B) = {\rm sgn} (B-A)$ when $|B|>|A|$)
\begin{eqnarray}
\bar{t} & = & \sqrt{\frac{B-A}{A+B}}\ \frac{1}{a}\ e^{aV} =
\sqrt{\frac{B-A}{A+B}}\ \frac{1}{a}\ e^{a(\eta + \xi)},\nonumber\\
\bar{z} & = & {\rm sgn} (A+B) \sqrt{B^{2}-A^{2}}\
\frac{1}{2a}\left(e^{aV} + e^{-aU}\right)\nonumber\\
 & = & {\rm sgn} (A+B) \frac{\sqrt{B^{2}-A^{2}}}{2a}\!\left(\!e^{a(\eta
  + \xi)} + e^{-a(\eta - \xi)}\!\right)\! .
\end{eqnarray}
It follows from this that in this wedge, we can write
\begin{eqnarray}
u & = & - \sqrt{\frac{B-A}{A+B}}\ \frac{1}{a}\ e^{-aU} = -
\sqrt{\frac{B-A}{A+B}}\ \frac{1}{a}\ e^{-a(\eta-\xi)},\nonumber\\
v & = & \sqrt{\frac{B-A}{A+B}}\ \frac{1}{a}\ e^{aV} =
\sqrt{\frac{B-A}{A+B}}\ \frac{1}{a}\ e^{a(\eta+\xi)},
\end{eqnarray}
so that we have 
\begin{eqnarray}
U & = & - \frac{1}{a}\ \ln \left(- \sqrt{\frac{A+B}{B-A}}\
au\right),\nonumber\\ 
V & = & \frac{1}{a}\ \ln \left(\sqrt{\frac{A+B}{B-A}}\
av\right).\label{rcoordinates}
\end{eqnarray}

On the other hand, in the other wedge labeled ``L'', we have
\begin{eqnarray}
\bar{t} & = & - \sqrt{\frac{B-A}{A+B}}\ \frac{1}{a}\ e^{aV} = -
\sqrt{\frac{B-A}{A+B}}\ \frac{1}{a}\ e^{a(\eta+\xi)},\nonumber\\
\bar{z} & = & - {\rm sgn} (A+B) \sqrt{B^{2}-A^{2}}\
\frac{1}{2a}\left(e^{aV} + e^{-aU}\right)\nonumber\\
& = &\!\! - {\rm sgn} (A+B)
\frac{\sqrt{B^{2}-A^{2}}}{2a}\!\left(\!e^{a(\eta+\xi)} 
+ e^{-a(\eta - \xi)}\!\!\right)\!\!,
\end{eqnarray}
so that we can write
\begin{eqnarray}
u & = & \sqrt{\frac{B-A}{A+B}}\ \frac{1}{a}\ e^{-aU} =
\sqrt{\frac{B-A}{A+B}}\ \frac{1}{a}\ e^{-a(\eta-\xi)},\nonumber\\
v & = & - \sqrt{\frac{B-A}{A+B}}\ \frac{1}{a}\ e^{aV} = -
\sqrt{\frac{B-A}{A+B}}\ \frac{1}{a}\ e^{a(\eta+\xi)}.
\end{eqnarray}
In the ``L'' wedge, therefore, we have 
\begin{eqnarray}
U & = & - \frac{1}{a}\ \ln \left(\sqrt{\frac{A+B}{B-A}}\
 au\right),\nonumber\\
V & = & \frac{1}{a}\ \ln \left(-\sqrt{\frac{A+B}{B-A}}\
av\right).\label{lcoordinates}
\end{eqnarray}

Let us next consider the quantization of a massless scalar field on
the light-front (equal $\bar{t}\ $) in the general light-front
frame. The field decomposition can be written as
\begin{eqnarray}
\phi (\bar{t},\bar{z}) & = & \frac{1}{\sqrt{2\pi}}\int_{0}^{\infty}
\frac{\mathrm{d}k}{2k}\big(e^{-i k\bar{t}} a_{1}
(k) + e^{-ik(\bar{t}-\frac{2}{A+B} \bar{z})} a_{2}
(k)\nonumber\\
 &  & \qquad\qquad\qquad\quad + {\rm Hermitian\ 
  conjugate}\big)\nonumber\\ 
 & = & \frac{1}{\sqrt{2\pi}}\int_{0}^{\infty}
\frac{\mathrm{d}k}{2k}\big(e^{-ik\bar{v}} a_{1}
(k) + e^{-ik\bar{u}} a_{2} (k)\nonumber\\
&  & \qquad\qquad\qquad\quad + {\rm
  Hermitian\  conjugate}\big),\label{glfexpansion}
\end{eqnarray}
where $a_{1} (k),a_{2} (k)$ can be thought of as the
annihilation operators for the null modes of the field components. The
field $\phi(\bar{t},\bar{z})$ can easily be checked to satisfy
light-front quantization conditions provided 
\begin{equation}
\left[a_{1} (k) , a_{1}^{\dagger} (k^{\prime})\right] =
2 k \delta (k- k^{\prime}) = \left[a_{2} (k),
  a_{2}^{\dagger} (k^{\prime})\right],
\end{equation}
and the vacuum of the theory satisfies
\begin{equation}
a_{1} (k) |0\rangle_{\rm GLF} = 0 = a_{2} (k)
|0\rangle_{\rm GLF}.\label{glfcondition}
\end{equation}

On the other hand, we can also quantize the theory in the two Rindler
wedges. Here, using the conformal coordinates, we can write the field
expansion as
\begin{eqnarray}
\phi(\eta,\xi) & = & \int_{0}^{\infty} \mathrm{d}K\Big(g_{K}^{\rm
  (R)} (U) b_{1} (K) + g_{K}^{\rm (R)} (V) b_{1} (-K)\nonumber\\
 &  & \qquad\qquad + g_{K}^{\rm
  (L)} (V) b_{2} (K) + g_{K}^{\rm (L)} (U) b_{2} (-K)\nonumber\\
 &  & \qquad\qquad\quad + {\rm Hermitian\
  conjugate}\Big),\label{rexpansion}
\end{eqnarray}
where we have defined the basis functions $g^{\rm (R)}, g^{\rm (L)}$
in the two wedges as
\begin{eqnarray}
g_{K}^{\rm (R)} (U) & = &\left\{ \begin{array}{cc}
\frac{e^{-iK(\eta-\xi)}}{\sqrt{2\pi}2K} =
\frac{e^{-iKU}}{\sqrt{2\pi}2K} & \text{in R},\\
0 & \text{in L},
\end{array}\right.\nonumber\\
g_{K}^{\rm (R)} (V) & = & \left\{\begin{array}{cc}
\frac{e^{-iK(\eta+\xi)}}{\sqrt{2\pi}2K} =
\frac{e^{-iKV}}{\sqrt{2\pi}2K} & \text{in R},\\
0 & \text{in L},
\end{array}\right.\nonumber\\
g_{K}^{\rm (L)} (U) & = & \left\{\begin{array}{cc}
0 & \text{in R},\\
\frac{e^{iK(\eta-\xi)}}{\sqrt{2\pi}2K} = \frac{e^{iKU}}{\sqrt{2\pi}2K}
& \text{in L},
\end{array}\right.\nonumber\\
g_{K}^{\rm (L)} (V) & = & \left\{\begin{array}{cc}
0 & \text{in R},\\
\frac{e^{iK(\eta+\xi)}}{\sqrt{2\pi}2K} =
\frac{e^{iKV}}{\sqrt{2\pi}2K} & \text{in L}.\label{basisfunctions}
\end{array}\right.
\end{eqnarray}
We can think of $b_{1} (K), b_{1} (-K)$ as the annihilation operators
for the two modes in the wedge ``R'' while $b_{2} (K), b_{2} (-K)$
correspond to the annihilation operators in the wedge ``L''. It is
easy to check that with the commutation relations
\begin{eqnarray}
\left[b_{1} (K), b_{1}^{\dagger} (K^{\prime})\right] & = & 2K \delta
(K-K^{\prime}) = \left[b_{1} (-K), b_{1}^{\dagger}
  (-K^{\prime})\right]\nonumber\\
\left[b_{2} (K), b_{2}^{\dagger} (K^{\prime})\right] & = & 2K \delta
(K-K^{\prime}) = \left[b_{2} (-K), b_{2}^{\dagger}
  (-K^{\prime})\right],\nonumber\\
 &  & \label{rcommutation}
\end{eqnarray}
the fields satisfy the conventional commutation relation for a theory
quantized on the light-front. The Rindler vacuum, which will be the
product of the vacua for the theories on the ``L'' and the ``R''
wedges satisfies
\begin{eqnarray}
b_{1} (K)|0\rangle_{\rm Rindler} & = & b_{1} (-K) |0\rangle_{\rm
 Rindler} =  0,\nonumber\\
b_{2} (K) |0\rangle_{\rm Rindler} & = & b_{2} (-K) |0\rangle_{\rm
 Rindler} = 0.\label{rcondition}
\end{eqnarray}

From the definition of the basis functions $g^{\rm (R)}, g^{\rm (L)}$
in (\ref{basisfunctions}), we see that they are not analytic and,
therefore, we cannot compare the Rindler vacuum to the GLF vacuum
directly. In fact, we note that while the annihilation operators in
(\ref{glfexpansion}) are the coefficients of positive frequency
eigenfunctions of the $\bar{P}_{0}$ operator (Hamiltonian), those in
(\ref{rexpansion}) correspond to coefficients of positive frequency
eigenfunctions of the boost operator $\bar{K}_{3}$ along $\bar{z}$ (We
note that
$\partial_{\eta} = a\left(\bar{t} \partial_{\bar{t}} + ((A+B)\bar{t} -
\bar{z})\partial_{\bar{z}}\right)$ is proportional to the boost
operator along the $\bar{z}$ axis). In order to compare the
two vacua, let us define a new set of basis functions (for the
expansion in the Rindler wedges) that are analytic,
\begin{eqnarray}
F_{K} (U) & = & \cosh \theta_{K}\ g_{K}^{\rm (R)} (U) + \sinh
\theta_{K}\ g_{K}^{{\rm (L)} *} (U),\nonumber\\
F_{K} (V) & = & \cosh \theta_{K}\ g_{K}^{\rm (R)} (V) + \sinh
\theta_{K}\ g_{K}^{{\rm (L)} *} (V),\nonumber\\
G_{K} (U) & = & \cosh \theta_{K}\ g_{K}^{\rm (L)} (U) + \sinh
\theta_{K}\ g_{K}^{{\rm (R)} *} (U),\nonumber\\
G_{K} (V) & = & \cosh \theta_{K}\ g_{K}^{\rm (L)} (V) + \sinh
\theta_{K}\ g_{K}^{{\rm (R)} *} (V),\label{newbasis}
\end{eqnarray}
with
\begin{equation}
\tanh \theta_{K} = e^{-\pi K/a}.\label{relation}
\end{equation}
It is easy to check that this new basis functions are analytic. For
example, we note that
\begin{eqnarray}
& & F_{K} (U) = \cosh\theta_{K}\left(g_{K}^{\rm (R)} (U) +
\tanh\theta_{K}\ g_{K}^{{\rm (L)} *} (U)\right)\nonumber\\
 &= &\!\! \frac{\cosh\theta_{K}}{\sqrt{2\pi}2K} \left\{\begin{array}{lc}
\!\! e^{-iKU} = \left(-\sqrt{\frac{A+B}{B-A}}\ au\right)^{iK/a} & \text{in
  R},\\
\!\! e^{-iK(U -i\pi/a)} = \left(-\sqrt{\frac{A+B}{B-A}}\ au\right)^{iK/a} &
\text{ in L},
\end{array}\right.\nonumber\\
 &  & 
\end{eqnarray}
and so on, where $u$ in the wedge ``L'' is assumed to lie slightly
above the principal branch of the logarithm. 

The field expansion in the two wedges in (\ref{rexpansion})
can now be expressed in terms of this basis function as
\begin{eqnarray}
\phi(\eta,\xi) & = & \int_{0}^{\infty} \mathrm{d}K\Big(F_{K} (U) c_{1}
(K) + F_{K} (V) c_{1} (-K)\nonumber\\
 &  & \qquad\qquad + G_{K} (V) c_{2} (K) + G_{K} (U) c_{2}
(-K)\nonumber\\
 &  & \qquad\qquad + \text{Hermitian
  conjugate}\Big),\label{newexpansion}
\end{eqnarray}
where we have defined new field operators resulting from the unitary
change in the basis as
\begin{eqnarray}
c_{1} (K) & = & \cosh\theta_{K}\ b_{1} (K) - \sinh\theta_{K}\
b_{2}^{\dagger} (-K),\nonumber\\
c_{1} (-K) & = & \cosh\theta_{K}\ b_{1} (-K) - \sinh\theta_{K}\
b_{2}^{\dagger} (K),\nonumber\\
c_{2} (K) & = & \cosh\theta_{K}\ b_{2} (K) - \sinh\theta_{K}\
b_{1}^{\dagger} (-K),\nonumber\\
c_{2} (-K) & = & \cosh\theta_{K}\ b_{2} (-K) - \sinh\theta_{K}\
b_{1}^{\dagger} (K).\label{operatortfn}
\end{eqnarray}
We note that the new creation and annihilation operators can be
easily seen to be related to the old ones through a Bogoliubov
transformation. Defining the formally unitary operator
\cite{das2,umezawa,israel1} 
\begin{equation}
U (\theta) = e^{-iG (\theta)},
\end{equation}
where
\begin{eqnarray}
G (\theta) & = &\! -i\! \int_{0}^{\infty} \!\!\mathrm{d}K
\frac{\theta_{K}}{2K}\Big(\big(b_{1} (K) b_{2} (-K) - b_{2}^{\dagger} (-K)
b_{1}^{\dagger} (K)\big)\nonumber\\
 &  & \qquad + \big(b_{1} (-K)b_{2} (K) - b_{2}^{\dagger}
(K)b_{1}^{\dagger} (-K)\big)\Big),
\end{eqnarray}
it is easy to check using the commutation relations
(\ref{rcommutation}) that we can write
\begin{eqnarray}
c_{1} (K) & = & U (\theta) b_{1} (K) U^{-1} (\theta),\nonumber\\
c_{1} (-K) & = & U (\theta) b_{1} (-K) U^{-1} (\theta),\nonumber\\
c_{2} (K) & = & U (\theta) b_{2} (K) U^{-1} (\theta),\nonumber\\
c_{2} (-K) & = & U (\theta) b_{2} (-K) U^{-1} (\theta).\label{bogoliubov}
\end{eqnarray}

We note that the basis functions $F_{K},G_{K}$ are positive frequency with
respect to GLF coordinates and hence the expansion in
(\ref{newexpansion}) can be directly
compared with the field expansion in (\ref{glfexpansion}). In
particular, we note that the GLF vacuum satisfying 
(\ref{glfcondition}) can also be written as
\begin{eqnarray}
c_{1} (K) |0\rangle_{\rm GLF} & = & c_{1} (-K) |0\rangle_{\rm GLF} =
 0,\nonumber\\
c_{2} (K) |0\rangle_{\rm GLF} & = & c_{2} (-K) |0\rangle_{\rm GLF} =
0.\label{newcondition}
\end{eqnarray}
On the other hand, using (\ref{bogoliubov}) as well as
(\ref{rcondition}), it follows now that we can relate the GLF vacuum
with the Rindler vacuum as
\begin{equation}
|0\rangle_{\rm GLF} = U (\theta) |0\rangle_{\rm Rindler}.
\end{equation}
Furthermore, from (\ref{operatortfn}) we note, for example, that
\begin{eqnarray}
c_{1} (K) |0\rangle_{\rm GLF} & = & \cosh\theta_{K}\!\left(b_{1} (K) -
e^{-\pi K/a} b_{2}^{\dagger} (-K)\right)\! |0\rangle_{\rm GLF}\nonumber\\
 & = & 0,
\end{eqnarray}
which implies that 
\begin{equation}
b_{1} (K) |0\rangle_{\rm GLF} = e^{-\pi K/a}
b_{2}^{\dagger} (-K) |0\rangle_{\rm GLF}.
\end{equation}
This shows \cite{das2,umezawa,israel1} that the Rindler observer
perceives the GLF vacuum as a
thermal vacuum at a temperature
\begin{equation}
T_{\eta\xi} = \frac{a}{2\pi}.
\end{equation}
The corresponding temperature in the GLF frame can then be obtained
from Tolman's law to be
\begin{eqnarray}
\frac{T_{\rm GLF}}{\sqrt{\bar{g}^{\rm (GLF)}_{00}}} & = &
\frac{T_{\eta\xi}}{\sqrt{\bar{g}^{\rm
      (conformal)}_{\eta\eta}}}\nonumber\\ 
{\rm or,}\quad T_{\rm GLF} & = & \sqrt{\frac{A+B}{B-A}}\
\frac{ae^{-a\xi}}{2\pi} = \sqrt{\frac{A+B}{B-A}}\ \frac{\alpha}{2\pi},
\end{eqnarray}
where we have used (\ref{acceleration}). This shows through a
systematic analysis from the Hilbert space point of view that a
uniformly 
accelerating observer would perceive the GLF vacuum to correspond to a
thermal vacuum with a temperature (in the GLF frame) given by
(\ref{tolman}). This is, of course, consistent with the results of the
earlier section, but as mentioned earlier is useful in showing that
any matrix element of the theory would appear as a thermal amplitude
to the accelerating observer.

\section{Summary}

In this paper, we have investigated the phenomenon of Unruh effect for
a massless 
scalar field theory quantized on the light-front in the general
light-front frame. In this case, there are two possible directions for
acceleration (as opposed to the Minkowski frame which is isotropic)
and we have determined the uniformly accelerating coordinates for both
the possible accelerations. By transforming the Green's function for
the massless scalar field quantized on the light-front to the
uniformly 
accelerating coordinate systems, we have shown that it has a thermal
character corresponding to a unique temperature given by Tolman's law
(\ref{tolman}) (independent of the direction of acceleration). We have
also carried out a systematic analysis of this phenomenon from the
point of view of the Hilbert space and have shown that a Rindler
observer finds the vacuum of the theory to correspond to a thermal
vacuum with the temperature given by Tolman's law, which in turn shows
that any amplitude of the theory would appear to be a thermal
amplitude to such an observer.

Finally, we note from the results obtained from our analysis that it
is an interesting question to determine whether
the vacuum of a quantum field theory quantized on equal time surface
is equivalent (through some Bogoliubov transformation or otherwise) to
that of the theory quantized on the light-front. {\em A priori} there
is no reason for such an equivalence, but the fact that {\em physical
amplitudes} in perturbation theory in the two theories agree on a case
by case basis, both at zero as well as finite temperature, (there is
no proof that this should happen in general and the agreement at zero
temperature depends crucially on the regularizations used since
the power counting arguments in the two theories are quite distinct)
makes it a worthwhile topic of study. Any direct relation between the
two vacua will lead to a better understanding of many aspects of both
(equal time and light-front) the theories.

\vskip .7cm

\noindent{\bf Acknowledgment}
\medskip

Two of us (A.D. and S.P.) would like to thank the Rockefeller
Foundation for support and Ms. Gianna Celli for kind hospitality at
the Bellagio Study Center where part of this work was done. This work
was supported in part by the US DOE Grant number DE-FG 02-91ER40685
and by CNPq as well as by FAPESP, Brazil.


\begin{thebibliography}{20}
\bibitem{typo} We take this opportunity to point out a couple of typos
  in reference [6]. Namely, the exponent in the propagator in eq. (38)
  should have a $\tilde{\bar{k}}$ and not $\bar{k}$ as it appears in
  the paper. Similarly, the last term on the right hand side of
  eq. (23) should have
  $\bar{\partial}_{3}\lambda$ instead of $\bar{\partial}_{3}\phi$.
\end{thebibliography}
\end{document}